# Self-Propelled Motion of a Droplet Induced by Marangoni-Driven Spreading

Yong-Jun Chen[1], Yuko Nagamine[2], Kenichi Yoshikawa[1,2] [†]

[1]Department of Physics, Graduate School of Science, Kyoto University, Oiwake-cho, Kitashirakawa, Sakyo-ku, Kyoto 606-8502, Japan

[2]Spatio-Temporal Order Project, ICORP, JST (Japan Science and Technology Agency), Kyoto 606-8502, Japan

[†]Corresponding author.

Tel: +81-75-753-3812, Fax: +81-75-753-3779,

E-mail: yoshikaw@scphys.kyoto-u.ac.jp

## Abstract

We report the generation of directed self-propelled motion of a droplet of aniline oil with a velocity on the order of centimeters per second on an aqueous phase. It is found that, depending on the initial conditions, the droplet shows either circular or beeline motion in a circular Petri dish. The motion of a droplet depends on volume of the droplet and concentration of solution. The velocity decreases when volume of the droplet and concentration of solution increase. Such unique motion is discussed in terms of Marangoni-driven spreading under chemical nonequilibrium. The simulation reproduces the mode of motion in a circular Petri dish.

PACS numbers: 47.20.Ma, 47.63.mf, 68.08.Bc

Introduction

Self-propelled motion is ubiquitous in biological world. Recently, there is intense scientific interest in self-propelled motion, such as macroscopic chemical Marangoni effect [1-4] and microscopic chemical powered transporter [5]. This is not only because of the purpose to get insight into the mechanism of biolocomotion but of the practical application of self-motion in robot design, Lab-on-a-Chip technique and microfluidic systems. The general mechanism of emergence of vectorial self-motion is the broken symmetry at front and rear. Particularly, Marangoni effect induces net slide of a droplet when wetting defect is formed [6]. Marangoni-driven motion has been discussed intensively in the previous publications [7-9]. An imbalance of surface tension pushes the droplet forward. Although spontaneous agitation at the interface induced by chemical Marangoni effect has been studied for more than a century, most reports have shown that the spontaneous motion is not regular with regard to both spatial and temporal changes. Recently, a certain kind of regular motion has been reported for an alcohol droplet [10]. However, the mechanism of such spontaneous motion has not been fully unraveled. In the present article, we report a unique mode of steady motion of a self-propelled droplet induced by Marangoni-driven speading.

Experiment

The experiment was performed using aniline oil ($\rho_{aniline}$=1.020±0.002 g/ml at 20 °C, 99.0%, Wako, Japan). Aqueous solutions containing various amounts of aniline were prepared. The aqueous solution was transferred to a circular glass Petri dish (diameter is 12 cm) and a rectangular vessel (10 cm×10 cm). The depth of the aqueous solution was from several centimeters to over 10 centimeters. A drop of aniline was placed on the air-solution interface through a pipette. The self-agitation of the droplet was monitored with a video camera and examined with image-analysis software. The measurements were carried out at room temperature (18±2 °C).

Results

Figure 1 exemplifies the self-propelled motion of an aniline droplet on an aqueous solution. After deposited on the shallow aqueous solution, the droplet quickly accelerated and achieved steady motion. When the volume of the droplet is large enough (for example, hundreds of microliters as in the case of Fig.1), the droplet moves continuously for hours. As shown in Fig. 1(a), a single droplet undergoes two different kinds of regular motion in a circular Petri dish depending on initial conditions (direction of motion and initial position): beeline motion and circular motion. The wall is apparently repulsive for the droplet as shown in the figure. When we deposit a droplet near the wall and induce the droplet along the wall gently, the droplet will move circularly along the circular wall, keeping a certain distance from the wall. The distance between the center of the droplet and the circular wall is almost constant during the circular motion. However, when the droplet is deposited far from the wall, the droplet will undergo beeline motion. The interaction between a droplet and the wall is subtle. A droplet in beeline motion prefers to switch to circular motion in a long term after bumps with the wall. Fig. 1(b) shows the spatio-temporal evolution of beeline motion between two parallel walls in a square vessel. The droplet undergoes to-and-fro motion between two walls. The wall repels the droplet, which decelerates when it approaches the wall. The repulsive force from the wall, as evaluated from the experimental tracking of the velocity of the droplet approaching the wall, is on the order of micro-Newtons. The droplet moves at a velocity of centimeters per second. The average velocity depends on the volume of the droplet and the concentration of the solution [Fig. 2(a)]. For the droplets with same volume, the velocity linearly depends on the concentration of aqueous solutions (Fig. 3).

Discussion

Aniline is partially miscible with water (concentration of aniline solution $c<3.6$vol% at $20\,°C$). Both dissolution through the droplet-solution interface and volatilization through the aniline-air interface are quite slow as observed in the experiment. The aniline spreads from the aniline droplet to air-solution interface and loses through air-solution interface mainly through volatilization and dissolution. We checked the Marangoni flow on air-solution interface around a droplet by putting a layer of hydrophobic powder on air-solution interface and the internal convection in the droplet by adding glass beads into the droplet. The Marangoni flow induced by surface-tension gradient is weak near the contact line of the self-propelled droplet and is not enough to supply driving force for the movement. The internal convection is not remarkable. On the other hand, after a minute amount of stearyl trimethyl ammonium chloride (surfactant) was added into aqueous solution, the motion ceased. This suggests that a change in interfacial tension is important for the occurrence of self-propelled motion. Thus, we will focus on the spreading process and the imbalance of surface tensions between the front and rear of a droplet. Figure 4(a) shows a schematic representation of the physico-chemical situation of an aniline droplet on an air-solution interface. The droplet sits and moves on the aqueous layer as spreading process proceeds (From the experimental observation, the spreading coefficient $S = \gamma_{a/s} - \gamma_{a/o} - \gamma_{o/s} > 0$, where $\gamma$ is the surface tension and the subscripts a, s and o represent air, aqueous solution and aniline, respectively). The contact line of the droplet is stable during motion when the concentration of solution c ≥ 2.8 vol%. Wave-propagation on the contact line is seen and motion of the droplet is complex when c < 2.8vol%. So we focus on the regular motion when c ≥ 2.8vol%. The droplet has a symmetric shape in beeline motion on a uniform solution in the front or back views [Fig. 4(b)]. But the difference in contact angles between the front and the rear is finite [Fig. 4(c)]. The wetting behavior of a liquid on a substrate is characterized by spreading coefficient $S$ and Hamaker constant $A$ [9, 12-13]. The positive spreading coefficient ($S > 0$) and the positive Hamaker constant ($A_{aos} > 0$ [14]) lead to the pseudo-partial wetting of aniline on the aqueous substrate [13]. In the case of pseudo-partial wetting, a macroscopic droplet with enough supply of liquid will coexist with an equilibrium thin film covering all over the substrate. However, in our experimental system, an equilibrium film can not be formed because of dissolution and volatilization at the air-solution interface. There is a precursor film with finite length in the vicinity of the contact line and the spreading process proceeds through this precursor film [Fig. 4(a)] [2]. The length of the precursor film depends on the rate of volatilization and dissolution on air-solution interface. At the front edge of the precursor film, a monolayer of aniline molecules spreads diffusively toward area with higher surface energy. The motion of a droplet persists for hours until the droplet consumes all its mass at the rate of a monolayer.

When the contact line is stable, the force balances on the contact line per unit length of the front and the rear are $\gamma_{a/o}\cos\alpha' + \gamma_{o/s}\cos\beta' = \gamma_{a/s}^{front}$ and $\gamma_{a/o}\cos\alpha + \gamma_{o/s}\cos\beta = \gamma_{a/s}^{rear}$, where contact angles $\alpha$, $\beta$, $\alpha'$, $\beta'$ are illustrated in Fig. 4(a). The values of $\alpha$ and $\alpha'$ are small, as illustrated in Fig. 4(d). Therefore, we obtain $\cos\alpha = \cos\alpha' \approx 1$. In addition, the relation of $\beta > \beta'$ is obtained [Fig. 4(c)]. According to the relations of force balance on the contact line mentioned above, we have $\gamma_{a/s}^{rear} < \gamma_{a/s}^{front}$. This indicates that the surface density of molecules at the rear should be higher than that at the front (aniline is a surface-active reagent) and thus a larger amount of aniline molecules spread toward the rear during the motion of the droplet. Using this model [Fig. 4(a)], the driving force per unit length is $f_D = \gamma_{o/s}(\cos\beta' - \cos\beta)$, where deformation of the air-solution interface is omitted. The driving force is characterized by geometric asymmetry. It has been shown that geometric asymmetry has possibility to induce driving force for self-propelled motion of a solid camphor scraping [15]. However, a liquid droplet can not sustain the difference of contact angles unless surface tensions at the rear and the front are different. The geometric asymmetry of the droplet actually is caused by a change of surface tension. When a droplet is static in a hole with pretty much the same size of the droplet, Marangoni vortexes were found near contact line of the droplet (see movie 3 [11]). Marangoni flow was induced by surface tension gradient in the vicinity of contact line. When the balance of surface tension was broken by perturbation, the droplet moves toward a direction and deforms at the rear. On an isotropic environment, the Marangoni flow on air-solution interface is principally symmetric around a static droplet. However, in the regular motion, fluid flow of solution, which passes the droplet, breaks the symmetry of Marangoni-driven spreading from the droplet. We attribute imbalance of surface tension to the effect of fluid flow of solution passing the droplet. As mentioned above, the perturbation initiates the asymmetry of surface tension around a droplet. The imbalance of surface tension causes the motion of the droplet. The motion induces the fluid flow of solution passing the droplet. The fluid flow of solution is cartooned in Fig. 4(f). The fluid flow contorts the way of Maragoni flow, which contains the surface-active molecules, as shown in Fig. 4(f). The motion itself breaks the symmetry of Maragoni-driven spreading around the droplet. The Maragoni flow points to the rear of the droplet. The aniline molecules spreading from lateral are taken to the rear by fluid flow of solution. The density of surface-active molecules is higher at the rear while the droplet meets the fresh surface of solution at the front. The asymmetric spreading of aniline molecules sustains the difference of surface tension at the rear and the front. The imbalance of surface tension is characterized by the difference of contact angles at the rear and the front. The difference of surface tensions at rear and front supplies the driving force for self-propelled motion. In beeline motion, when viscous drag $F_{vd}$ acting on a droplet balances driving force $F_D$ from surface-tension difference at front and rear, the droplet moves steadily. We deduced steady velocity of beeline motion by the balance of viscous drag and driving force $F_{vd} = F_D$ (see Appendix). The predicted velocity is shown in Fig. 2(b) and agrees well with experimental measurements in Fig. 2(a).

Our experiment indicates that a wall repels a droplet. The repulsive force includes two parts. On the one hand, the wall impedes the further diffusive spreading of surface-active molecules and this leads to a lower surface tension near the wall (Wall effect). On the other, the meniscus formed near the wall induces a repulsive net force. In circular motion, wall effect and meniscus effect supply centripetal force. To get insight into the wall effect, let us compare the repulsive net force from meniscus effect with centripetal force in circular motion. As mentioned before, a droplet keeps a certain distance from a wall when it moves circularly in a circular Petri dish. The space between the droplet and the wall [Fig. 6(a) and (b)], $x$, is in the range of 0.6-1.2 cm in circular motion when 2.8 vol% $\leq c \leq$ 3.3 vol% in a circular Petri dish (detail not shown). Centripetal force $f_c = \rho_{aniline} V U^2 / R_0$ ($V$ is volume of a droplet and $R_0 = 5.0$ cm is radius of circular motion) is plotted in Fig. 6(c) according to experimental data of velocities $U$ in a circular Petri dish [Fig. 2(a)]. As for the meniscus effect, the net force per unit length due to the deformation of meniscus near the wall can be expressed as $f_m = \gamma_{a/s} - \gamma'_{a/s} \cos\theta$, where $\gamma'_{a/s}$ and $\theta$ are shown in Fig. 6(a). Assume total wetting of the solution on the glass wall. $\cos\theta$ is determined by the shape of the meniscus [9]: $(x - x_0)/\kappa^{-1} = \cosh^{-1}[\sqrt{2}/(1-\cos\theta)^{1/2}] - (1+\cos\theta)^{1/2}$ ($\theta = 90°$ when $x = 0$, $x_0$ is constant), where $\kappa^{-1} = \sqrt{\gamma_{a/s}/\rho g}$ is the capillary length and g is gravity acceleration ($\gamma_{a/s} = 49$ mN/m from experimental measurement when c=2.8 vol%, and thus $\kappa^{-1}=0.21\,\text{cm}$). By setting $\gamma'_{a/s}=\gamma_{a/s}$ for simplicity, the net force per centimeter due to meniscus effect is shown in Fig. 6(d). When $x>0.6\,\text{cm}$, the net force per centimeter due to meniscus effect is less than 0.5 µN when c=2.8 vol%. The net force from the meniscus effect is not enough (when $x>0.6$ cm) to supply centripetal force in circular motion (diameter of droplets is less than 2cm) as shown by comparison of the results in Fig. 6(c) and (d). This indicates that the wall effect, which results in a lower surface tension near the wall, plays a role in centripetal force of the circular motion. But as shown in Fig. 6(d), net force due to meniscus effect will increase fast when x decreases at $x<0.6$ cm. Thus, when a droplet approaches the wall, it will be repelled by meniscus effect. The repulsion from the wall suggests a way to control the motion of a droplet. We put a gap on the path of motion. The repulsive force from the glass wall deforms the droplet and then droplet passes through the gap narrower than the normal size of the droplet [Fig. 7]. The aniline droplet does not prefer to attach on the glass wall. The imbalance of surface tension at rear and front pushes the droplet forward in the gap.

Numerical simulation

We now discuss the mechanism of mode-selection in a circular Petri dish as shown in Fig. 1(a). The active motion is driven by dissipative process. A droplet prefers to maintain a steady velocity $U$. The net force on a droplet is related to the balance between viscous drag and difference of surface tensions at rear and front. When the velocity of a droplet is smaller than that of the steady state, the net force dominated by the imbalance of surface tensions at rear and front will accelerate the droplet. But when the velocity of a droplet is larger than that of the steady state, the net force dominated by the viscous drag will decelerate the droplet. The motion of our droplet is similar to that of a biological object or an active particle with active friction. Langevin dynamics has been used to describe motion of Brownian particle with energy depot [18]. In our system, the persistent Marangoni-driven spreading from droplet to aqueous phase supplies kinetic energy for self-propelled motion. Stochastic force is neglected. Including active friction from Marangoni-driven spreading and viscous drag, we write governing equation as Langevin dynamics without stochastic

term:

$$\frac{d\bar{U}'}{dt} = -\xi(|\bar{U}'|^2)\bar{U}' - \nabla\varphi \qquad (1)$$

According to the expression of viscous drag, we write the velocity-dependent active friction coefficient as $\xi(|\bar{U}'|^2) = \mu(|\bar{U}'|^2 - U^2)$. $\varphi$, $\bar{U}'$, $|\bar{U}'|$, $U$ and $\mu$ are the effective potential from the repulsion of the wall, velocity, absolute velocity, steady velocity and a constant, respectively. Here we use velocity-dependent friction coefficient $\xi(|\bar{U}'|^2)$ to characterize the balance of driving force and viscous drag. The positive or negative value of $\xi(|\bar{U}'|^2)$ indicates the process of deceleration or acceleration. The active friction in Eq. 1 would like to maintain steady velocity of motion [19]. This corresponds to the motion of our system. To determine the effective potential from the wall, we tracked the motion of a droplet between two parallel walls in a rectangular vessel as shown in Fig. 1(b). The net force was estimated according to the acceleration of the droplet (Fig. 8). We constructed a parabolic effective potential $\varphi$ from the wall according to the spatial evolution of the net force in Fig. 8. The effective potential includes wall effect and meniscus effect. We choose $\varphi$ as

$$\varphi(a) = \begin{cases} -9\sigma & (0 \le a \le a_0) \\ \sigma(a^2 - 6a) & (a \ge a_0) \end{cases}$$

where $a = (x^2 + y^2)^{1/2}$ is the radial distance to the center of the circular Petri dish in an xy orthogonal coordinate system (Fig. 9), $a_0 = 3$ and $\sigma$ is an adjustable constant. The constants $2\sigma$ and $a_0$ characterize gradient of force from wall and scope where effect of wall works. The value of $a_0$, which depends on radius of Petri dish, does not have much influence on results because we do not fix position of wall in this model. Steady velocity is determined by concentration of solution and size of a droplet. Figure 9 shows the numerical results for the motion after a minute perturbation under specific initial conditions in a circular Petri dish. At the beginning, the droplet accelerates quickly as that in the experiment. The wall reflects the droplet when the droplet approaches the wall. The mode of motion depends on the initial position and the droplet tends to approach its steady state as shown by the time evolution of the velocity in Fig. 9. We found circular mode of motion when a droplet has an initial velocity in Fig. 10. In the simulation, steady motion is not always circular motion as shown in the Fig. 9 & Fig. 10. The initial condition determines the subsequent mode of motion. From different initial positions, droplets undergo different modes of motion. The behavior agrees with the experimental observation [Fig. 11]. But, in the experimental case, the freedom of rotation of the droplet subtly counteracts the repulsive effect from the wall and leads to smoother motion. A droplet prefers to switch from non-circular motion to circular motion through repulsive potential and then be self-confined near the wall as shown in Fig. 11(b).

As shown in the experimental observation, steady velocity depends on the size of a droplet and the concentration of solution [Fig. 2(a)]. The size and concentration of solution have important effect on the motion of a droplet. For a droplet on a solution with specific concentration, there is a characteristic steady velocity. Figure 12 shows the numerical results of motion with various steady velocities $U$ under certain initial condition. When steady velocity $U = 0$ (This corresponds to the motion on a saturated solution, we set effective potential in Eq.2 to zero), the viscous drag will decelerate the droplet and the velocity comes near zero soon. For non-zero steady velocity, the mode of motion depends on the steady velocity, that is, droplets with different size have different modes of motion under same initial condition. Particularly, we can not find a circular mode of motion when steady velocity $U < 0.5$. It indicates that a very big droplet, which moves slowly in a circular Petri dish, can not undergo circular motion as found in the experiment. But here it is important to point out that we use the same effective potential field in simulations of motion with various steady velocities. Actually, the effective potential field for a droplet from wall effect should depend on size of the droplet and concentration of solution. And another point is that we do not fix the position where the wall is located. In our experimental system, wall effect and meniscus effect work only in a very limited space near the wall. Thus, when the real position of the wall is considered, a droplet with a large velocity will be reflected by meniscus and wall itself directly while a droplet with a small velocity should be sensitive to the effective potential. This is the case we have observed in the experiment. As a remark for the simulation, the size of the droplet and the inertial effect are not included in the model for simulation. Principally, the droplet is a three-dimensional object and the net force is determined by an integral around droplet. For a big droplet (For example, with diameter of 1-2cm), size of the droplet and inertial effect have important effect on the mode of motion. The big droplet will penetrate the effective potential field by inertial effect and bump with the wall.

Conclusion

In summary, we have demonstrated the unique motion of a droplet. In contrast to previous studies, we found that droplets exhibit repulsive interaction between each other and with a wall. This indicates the possibility to control an individual droplet. Also the present system is suitable for observing swarming phenomenon in an easy-to-do experiment with a good reproducibility. Further studies along these lines may be promising.

Acknowledgement

We are grateful to Ken Nagai, Shun Watanabe and Dr. Nobuyuki Magome for providing technical assistance, and also thank Dr. Takashi Taniguchi, Yutaka Sumino and Professor David Quéré for their helpful discussions. We would like to thank anonymous reviewers for helpful comments on the improvement of the paper. Y. J. C thanks Ichikawa foundation for financial support. This work was supported by a Grant-in-aid for Creative Scientific Research (Project No. 18GS0421).

Appendix: Predicted velocity

Here, we discuss the steady velocity by the balance of viscous drag $F_{vd}$ and driving force $F_D$. Generally, the driving force can be written as

$$F_D = \oint_l \gamma_{o/s} \cos\beta_c \hat{r} \cdot \bar{n} \, dl$$

where $l$ is contact line, $\hat{r}$ is unit vector in the direction of $\bar{\gamma}_{a/s}$, $\bar{n}$ is unit vector in the direction of motion and $\beta_c$ is contact angle of droplet-solution interface with air-solution interface. We estimated Reynolds number (R) as $40 < R = U\sqrt{B}/\nu < 200$ when concentrations of solutions are 2.8 vol%, 3.0 vol% and 3.3 vol%, according to Fig. 2(a) and Fig.6(a), where $\nu = 1.0 \times 10^{-2}$ cm$^2$/s is kinetic viscosity of water and B is arc reference area perpendicular to the direction of motion [Fig. 4(e)] (characteristic length of the reference area is defined as $\sqrt{B}$): $B = (1/4h^2)[(h^2+r^2)^2 \sin^{-1}(2rh/(h^2+r^2)) - 2rh(r^2-h^2)]$ (r is radius of a droplet and h is depth of a droplet immersed in the solution). And viscous drag acting on a droplet is $F_{vd} = 1/2\rho U^2 C_d B$, where drag coefficient $C_d = 24f(R)/R$, and $f(R) = 1 + 0.15R^{0.687} \approx 0.41R^{0.519}$ when Reynolds number is $40 < R < 200$ [16], and $\rho$ is density of the aqueous solution. The droplet moves steadily when $F_D = F_{vd}$. It is difficult to estimate the distribution of the contact angle $\beta_c$ on the contact line. By setting $F_D = 2\gamma_{o/s} r(\cos\beta' - \cos\beta)$, we obtain

$$U = \left[ \frac{\gamma_{o/s} r}{2.46 \times \rho B^{0.760} \nu^{0.481}} (\cos\beta' - \cos\beta) \right]^{0.658} \quad (A1)$$

Contact angles $\beta$ and $\beta'$, diameters D=2r and depths h of droplets on various solutions are measured experimentally and shown in Fig. 5. We use $\rho = 1.0$ g/cm$^3$ and $\nu = 1.0 \times 10^{-2}$ cm$^2$/s. The surface tension of droplet-solution interface $\gamma_{o/s}$ is shown as the inset in Fig. 2(b). Equation A1 does not include the volume of droplets directly. The diameter D, depth h, and contact angles are functions of volume of the droplet as shown in the experimental measurements [Fig. 5]. To get specific dependency of parameters in Eq. A1 on volume of the droplet and concentration of solution is difficult. The contact angles are affected by concentration of solution [Fig. 5(b)], which leads to concentration-dependent motion of a droplet [see Fig. 3]. Thus, we obtain the evolution of velocity as a function of volume by substituting to Eq. A1 parameters of r, h, $\beta$ and $\beta'$, which are corresponding to each value of the volume in Fig. 5. The log-log plot of the predicted steady velocities against the volume of droplets in beeline motion is shown in Fig. 2(b) [17]. The velocity against concentration of solution is plotted in Fig. 3. Comparing with experimental results, the calculated results based on Eq. A1 are reasonable (Fig. 3). The slope $\eta$ of the fitting lines in Fig. 2(b) agree with that in Fig. 2(a), suggesting that the predicted results agree well with that of the experimental observations. The deviation of predicted results from experimental data is mainly due to error of estimation of viscous drag and too simple consideration of the distribution of contact angles on the contact line.

---

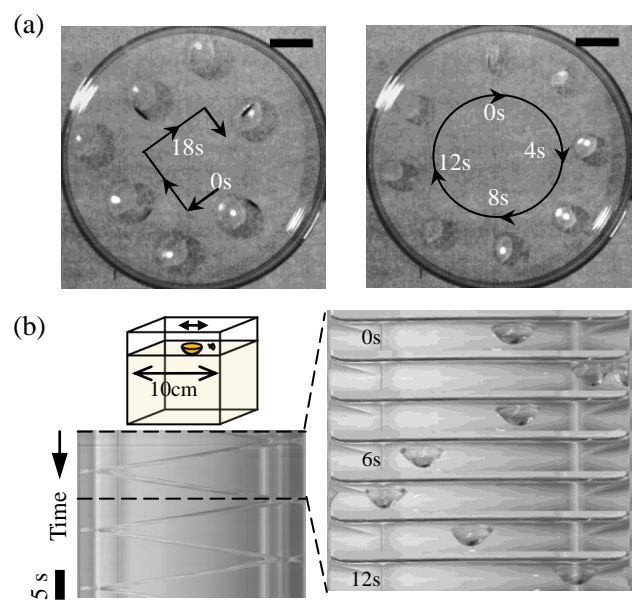

FIG. 1 (color online) Self-propelled motion of droplets. (a) Spatio-temporal evolution of beeline motion (left) and circular motion (right) in a circular Petri dish. The droplets have volumes of 975 μl (left) and 325 μl (right), respectively. (b) Spatio-temporal image of beeline motion between two parallel walls in a square vessel. The concentration of the solution in (a) and (b) is 2.8 vol%. Scale bars are 2 cm. As for detail motion, see movie 1 for circular motion and movie 2 for beeline motion [11].

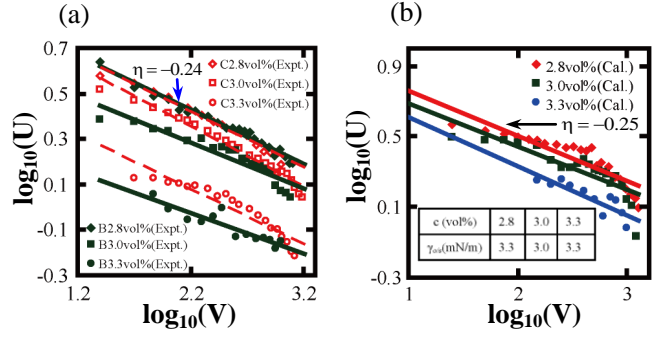

FIG. 2 (color online) $Log_{10}$-$log_{10}$ plot of velocity U (unit: cm/s) and volume of droplet V (unit: μl) of motion on various solutions. (a) Experimental results (Expt.). Velocity is average velocity. B: beeline motion in a square vessel, C: circular motion in a circular Petri dish. The droplets move on 500 ml of solution in a square vessel (Beeline motion) and 200 ml of solution in a circular Petri dish (Circular motion), respectively. (b) Calculated steady velocity of beeline motion (Cal.). The inset shows the interfacial tension at the droplet-solution interface ($\gamma_{o/s}$) measured by the Wilhelmy plate method at room temperature. The lines in the figures are fitting to the data and $\eta$ is slope of fitting lines.

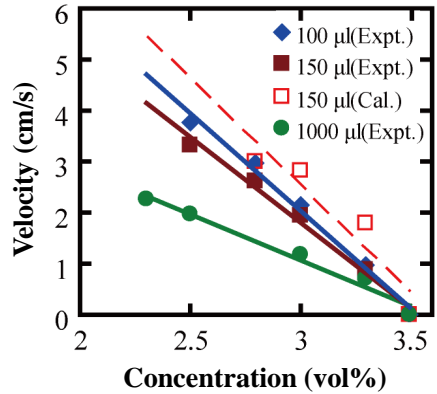

FIG. 3 (color online) Velocity depending on concentration of solution. The volumes of droplets are 100μl, 150μl and 1000μl, respectively. The comparison of Experimental data (Expt.) and calculated result (Cal.) is presented (V=150μl). The lines is the linear fitting to the data.

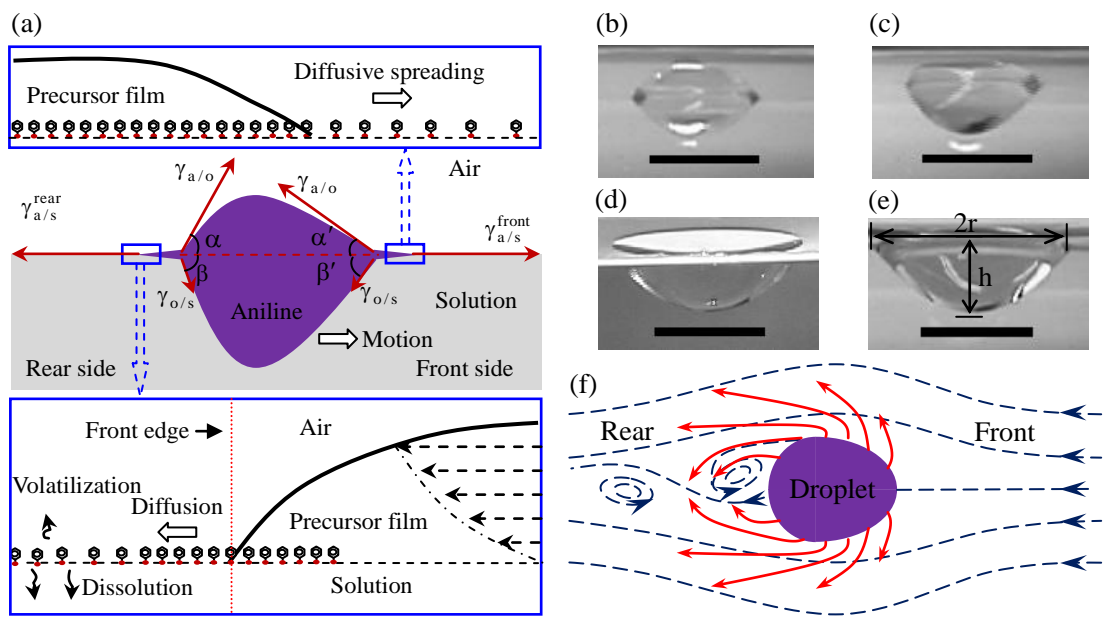

FIG. 4 (color online) Scenario of a self-propelled droplet on an air-solution interface. (a) Precursor film model of a droplet from which aniline spreads on an aqueous solution. (b) Side view of a droplet in beeline motion (front or back view). (c) and (d) Side views of the lower (c) and flat upper (d) parts of droplets in beeline motion on 2.8 vol% solution. (e) Side view of reference plane perpendicular to the direction of motion. Scale bars are 1cm. (f) Cartoon of fluid flow (dashed line) and Marangoni flow (solid line) near a droplet in motion. The Reynolds number is from 40 to 200. The fluid flow is not only layer flow but "Karman vortex street" on the rear. The Marangoni flow is distorted by fluid flow.

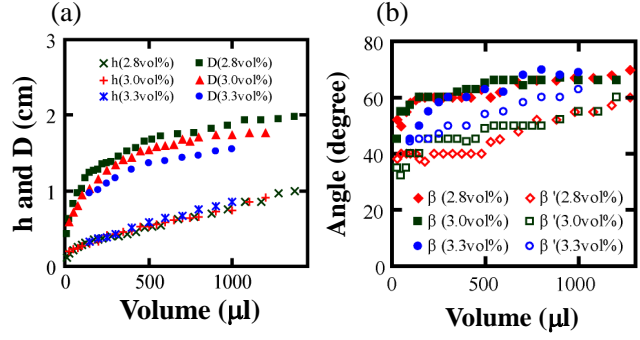

FIG. 5 (color online) Geometric parameters of droplets in beeline motion on various solutions in a square vessel measured in experiments. (a) Depths (h) and diameters (D=2r) of droplets on various solutions in a square vessel. (b) Contact angles ($\beta, \beta'$) on front and rear.

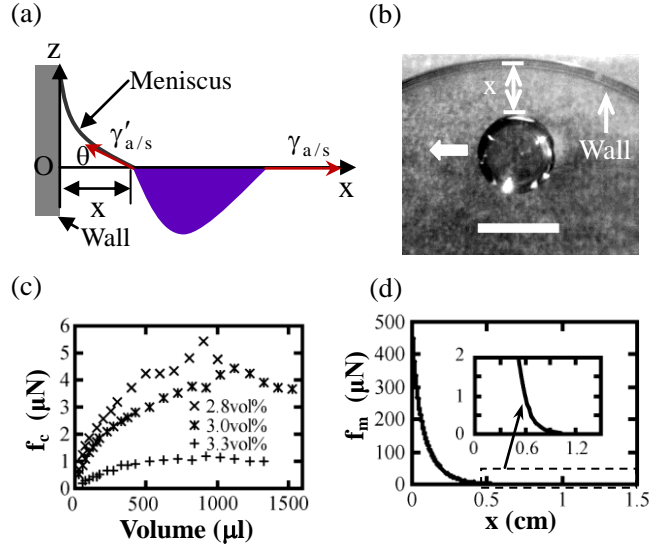

FIG. 6 (color online) Interaction of the droplet with the wall. (a) A schematic of a droplet near a wall (side view). $\theta$ is the angle between tangent of meniscus and horizontal. (b) Top view of a droplet moving near a wall. Scale bar is 2cm. (c) Calculated centripetal force of circular motion in a circular Petri dish. (d) Net force due to Meniscus effect. Demonstrated is net force per centimeter from the meniscus effect when c=2.8 vol%.

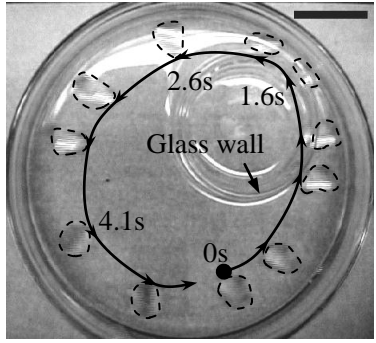

FIG. 7 (color online) A droplet passes through a narrow gap. Dashed lines on the droplet show shape of the droplet. Scale bar is 2cm. As for the detail of the motion, see the movie 4 and movie 5 [11].

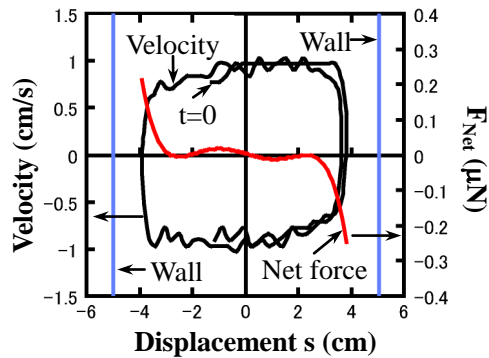

FIG. 8 (color online) Tracking of beeline motion between two parallel walls in Fig. 1(b). Plotted are spatio-temporal traces of velocity and net force ($F_{Net}$) experienced by a droplet at the air-solution interface (volume of droplet: 200 μl, concentration of solution: 3.3 vol%) which underwent the motion in Fig. 1(b). s is the displacement of the droplet. We estimated net force from the time evolution of velocity $F_{Net} = \rho V\, dU/dt$. The positions of wall are shown in the figure.

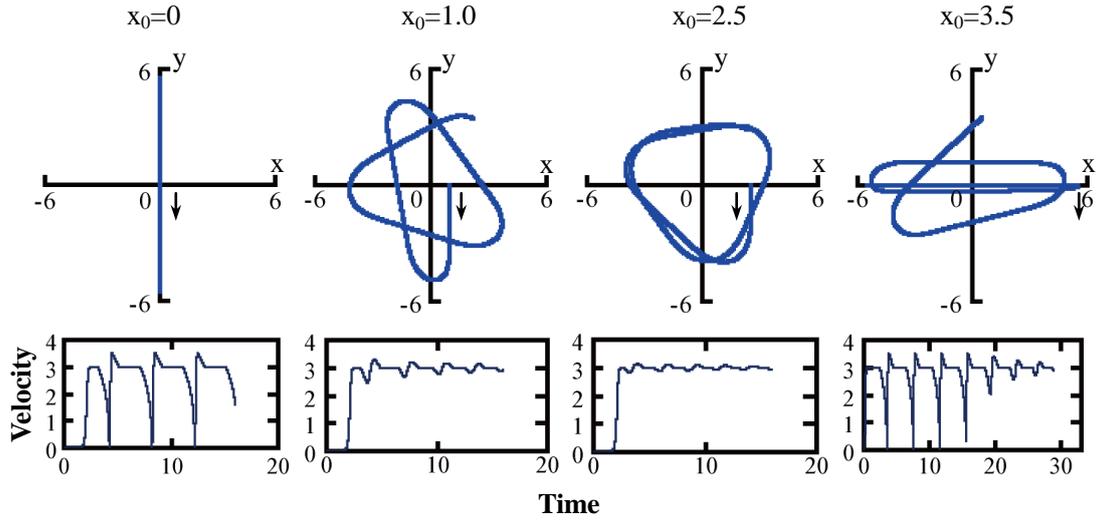

FIG. 9 (color online) Numerical results of motion after a minute initial perturbation in a circular Petri dish. Upper: trajectories of motion, Lower: time traces of velocities. A droplet chooses different modes depending on initial position $x_0$. Initial conditions: $y_0=0$, $U_x^0=0$ and $U_y^0=0.00000001$. Other parameters: $\mu=1$, $\sigma=3$. The arrows show the direction of motion from initial positions.

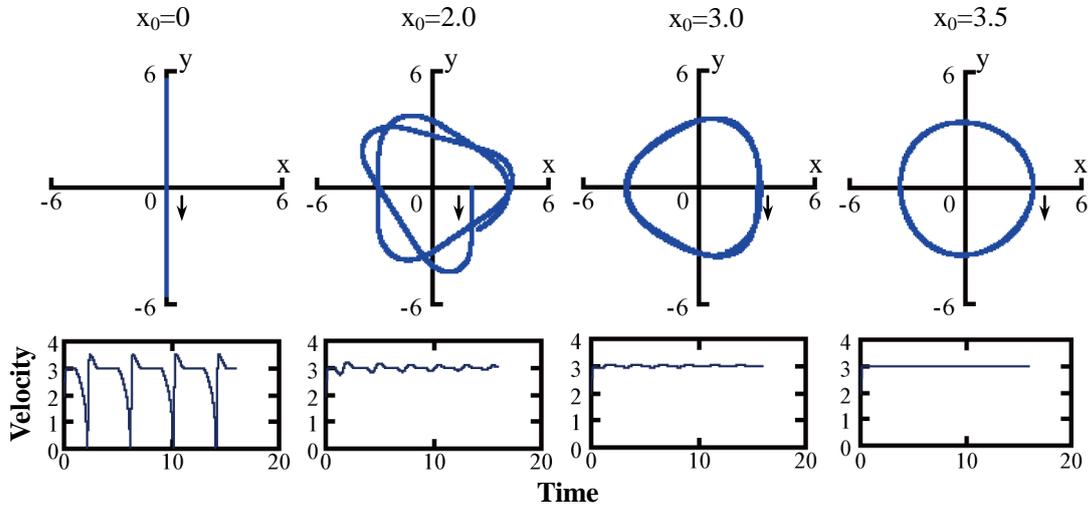

FIG. 10 (color online) Numerical results of motion with a non-zero initial velocity in a circular Petri dish. Upper: trajectories of motion, Lower: time traces of velocity. A droplet chooses different modes depending on initial position $x_0$ shown in the figures. Initial conditions: $y_0=0$, $U_x^0=0$, and $U_y^0=2$. Other parameters: $\mu=1$, $\sigma=3$. The arrows show the direction of motion from initial positions.

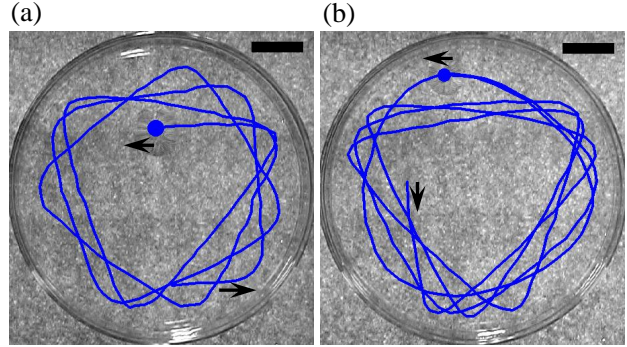

FIG. 11 (color online) Experimental tracking of motion in a circular Petri dish. (a) A typical trajectory of beeline motion. Volume of the droplet is 1000 μl. (b) Mode switching from beeline motion to circular motion. Volume of the droplet is 400 μl. Scale bars in the figures are 2 cm. The concentration of solution is 2.8 vol%. The arrows show the direction of motion. See movie 6 for (a) and movie 7 for (b) [11].

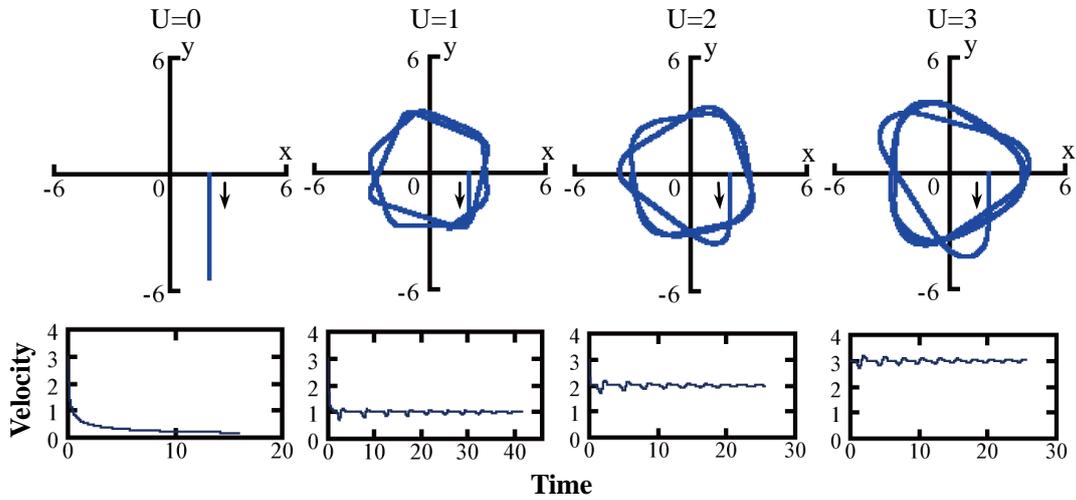

FIG. 12 (color online) Numerical results of motion with various steady velocities. Upper: trajectories of motion, Lower: time traces of velocities. A droplet chooses different modes depending on steady velocity $U$. Initial conditions: $x_0=2$, $y_0=0$, $U_x^0 = 0$, and $U_y^0 = 3$. Other parameters: $\mu = 1$, $\sigma = 0$ for $U = 0$ and $\sigma = 3$ for $U \neq 0$. The arrows show the direction of motion from initial positions.